\documentstyle[12pt]{article}

\author{V.A.Tsokur and Yu.M.Zinoviev
       \thanks{E-mail address: ZINOVIEV@MX.IHEP.SU} \\
        {\it Institute for High Energy Physics} \\
        {\it Protvino, Moscow Region, 142284, Russia}}
\title{$N=2$ Supergravity Models\\
        Based on the Nonsymmetric Quaternionic Manifolds \\
       I. Symmetries and Lagrangians}
\date{March 1996}

\begin{document}

\maketitle

\begin{abstract}
  In this paper we consider the $N=2$ supergravity models in which
the hypermultiplets realize the nonlinear $\sigma$-models,
corresponding to the nonsymmetric (but homogeneous) quaternionic
manifolds. By exploiting the isometries of appropriate manifolds we
give an explicit construction for the Lagrangians and
supertransformation laws in terms of usual hypermultiplets in the
form suitable for the investigation of general properties of such
models as well as for the studying of concrete models.
\end{abstract}

\newpage

\section*{Introduction}

  Recently there an interest in the theories with $N=2$ supersymmetry
has revived. In the stringy context the reason was that one could use
the same Calabi-Yao manifolds for compactification of heterotic
string (obtaining the theory with $N=1$ supersymmetry) and for the
type II string thus having $N=2$ supersymmetry. The latter leads to
the restrictions on the geometrical properties of such models like
moduli spaces and so on. Another reason is related with
quantum properties of the theories with $N=2$ supersymmetry giving
a possibility to study nonperturbative phenomena \cite{Sei94}.
From the phenomenological point of view models with extended
supersymmetries do not look promising due to severe problems arising
in any attempts to construct even semirealistic theory. Generally,
one faces three kinds of problems:
\begin{itemize}
\item  Spontaneous supersymmetry breaking. \\
Any realistic model should rely on the mechanism of spontaneous
supersymmetry breaking which gives two essentially different scales,
in this the cosmological term must be automatically equal to zero for
any values of these parameters. Two scales of breaking are necessary
because the breaking with only one scale (when two gravitini remain
mass degenerate) leaves the theory vectorlike \cite{Cre85,Ito87}. In
order to be able to reproduce at low energies the "standard" $N=1$
supersymmetric phenomenology, such a model should admit, as a
particular case, the partial super-Higgs effect then the $N=1$
supersymmetry remains unbroken and the corresponding gravitino ---
massless. It turns out that such a breaking is indeed possible
\cite{Cec86a,Zin86}, moreover it has been shown \cite{Zin90} that
there exist three different hidden sectors having desirable
properties. This, in turn, allowed one to consider the spontaneous
supersymmetry breaking in a wide class of$N=2$ supergravity models
and calculate the soft breaking terms that arose after the breaking
had taken place \cite{Zin86a,Zin90}.
\item  Gauge symmetry breaking. \\
Generally, the scalar field potential in $N=2$ gauge theories as well
as in $N=2$ supergravity models has quite a lot of flat directions
giving one a possibility to introduce non-zero vacuum expectation
values for the appropriate scalar fields thus breaking the gauge
symmetry spontaneously. However, after the spontaneous supersymmetry
breaking at least part of these flat directions is lost. For
example, one of the general properties of the mechanism of
spontaneous supersymmetry breaking described above is that the scalar
fields from the vector multiplets unavoidably acquire masses. Let us
stress, however, that analogous problem (the positive mass square for
the Higgs field) appears in the $N=1$ theories as well.
\item  Fermionic mass spectrum. \\
In the theories with $N=2$ supersymmetry this problem turns out to be
much more difficult. First of all, as we have already mentioned, with
the unbroken supersymmetry or even when there is only one breaking
scale, the theory is vectorlike and one has to have spontaneous
supersymmetry breaking with two different scales before trying to
lift the mass degeneracy between usual and mirror fermions. But even
after the problem with the supersymmetry breaking has been solved,
one has to have an appropriate set of Yukawa couplings to generate
correct fermionic mass spectrum. As is well known, in $N=2$ gauge
theories there exists only one type of such couplings, where the
scalar fields come from the vector multiplets. In the "minimal"
coupling of these theories to $N=2$ supergravity (based on the
quaternionic spaces that are symmetric ones) the situation remains to
be essentially the same --- there are no Yukawa couplings between
scalar and spinor fields of the hypermultiplets.
\end{itemize}

  However, there exist quaternionic manifolds that are not
symmetric manifolds (but they are homogeneous ones which is important
for the possibility for such models arise in superstrings). For the
first time the classification of these spaces has been given by
Alekseevsky \cite{Ale75}, later on the properties of these manifolds,
especially their symmetry properties, were studied in a number of
papers \cite{Cec89,Wit90,Wit93,Wit94,Wit95}. Two properties will
be very important for us in this work:
\begin{itemize}
\item As we will show below, all these models contain as a universal
part one of the three hidden sectors \cite{Zin90} (corresponding to
the nonlinear $\sigma$-model $O(4,4)/O(4)\otimes O(4)$) admitting a
spontaneous supersymmetry breaking with two arbitrary scales and
without a cosmological term.
\item In the most general case (we will denote such models $V(p,q)$,
see below) there exist cubic invariants (due to the fact that part of
the hypermultiplets transformed under the vector representation of
group $O(p)$, while others --- under the spinor ones) related to the
appropriate $\gamma$-matrices. This leads to a possibility for
Yukawa couplings to be generated.
\end{itemize}

  The Alekseevsky classification \cite{Ale75} for the quaternionic
manifolds is reduced to the classification of admittable quaternionic
algebras with the dimension (always multiple of four) equal to the
dimension of the manifold. This means, that each generator in the
algebra corresponds to the physical scalar field in the
hypermultiplets being the coordinate of the manifold and playing the
role of the Goldstone one for this transformation. In the cases when
the quaternionic manifold is symmetric, the isometry algebra turns
out to be larger (e.g. $Sp(2,2n)$, $SU(2,n)$ or $O(4,n)$ for the well
known cases). But even in general cases the isometry algebra of the
quaternionic manifold appears to be larger than the minimal one
arising in classification \cite{Wit90}-\cite{Wit95}. This
allows one by using the invariance under these algebras to construct
and investigate the appropriate nonlinear $\sigma$-models. In
particular, we will see that in all cases it is possible to have
linearly realized $SU(2)$ subgroup, corresponding to the natural
symmetry of the $N=2$ superalgebra. 

   All generators in the quaternionic algebra (apart from the ones in
hidden sector) come in three types which following Alekseevsky we
will denote as $X$, $Y$ and $Z$. In this, there exist two general
series of models.%
\footnote{Really, there exist more models, as it has been shown in
\cite{Wit93,Wit94}, but in this paper we restrict ourselves to only
two general types of models}
In the first one, that we will denote $W(p,q)$, the $X$-type
hypermultiplets are absent, while the number $p$ of $Y$-type
hypermultiplets and the number $q$ of $Z$-type ones are arbitrary. In
the second one denoted as $V(p,q)$ there exist $p$ sets of $X$-type
hypermultiplets while the number of $Y$ and $Z$ hypermultiplets are
equal to $q d(p)$, where $d(p)$ --- dimension of spinor
representation of $O(p)$ group. In the four special cases ($q = 1$
and $p=1,2,4,8$) these manifolds turn out to be symmetric with the
isometry algebras being $F_4$, $E_6$, $E_7$ and $E_8$. Thus, the
minimal model that contains all three types of hypermultiplets is the
$F_4$ one and we will use this model as our starting point. Namely,
in the following section we will show how one can construct the
linear combinations of the $F_4$ generators corresponding to the
generators in the Alekseevsky classification. Moreover, we will
explicitly construct the subalgebra of the whole $F_4$ algebra which
admits natural generalization to the case of nonsymmetric
quaternionic manifolds.

  In the next section we consider the $W(p,q)$ models for arbitrary
$p$ and $q$. As is known \cite{Cec89}, in the partial case $q=0$ this
model coincides with the well known $O(4,4+p)/O(4)\otimes O(4+p)$
one. This allows one to have a very simple realization of this model
which could be easily generalized to the case of arbitrary $q$.

  Later, as a preliminary step to the construction of the general
$V(p,q)$ models, we consider one more partial case --- $q=0$, i.e.
the case when only $X$-type hypermultiplets are present. This model
also corresponds to the similar  $O(4,4+p)/O(4)\otimes O(4+p)$
model but in different parameterization. Using this fact, we have
managed to construct the realization with the correct isometry
algebra (i.e. global symmetry of the bosonic Lagrangian).

  Both models $W(p,q)$ and $V(p,0)$ contain the same hidden sector,
corresponding to the non-linear $\sigma$-model $O(4,4)/O(4)\otimes
O(4)$ but in the very different parameterizations. So, to join these
models into the general $V(p,q)$ one we have to make a reduction for
both of them in order to bring the hidden sectors to the similar
form. Unfortunately, this enlarges the number of fields in terms of
which the model is described and makes all the formulas rather long.
Nevertheless, we have managed to construct an explicit Lagrangian
invariant under the local $N=2$ supertransformations and global
bosonic transformations, corresponding to the appropriate
quaternionic algebras.

\section{$F_4$-model}

   As we have already mentioned in Introduction, the minimal
model that contains all necessary ingredients for the construction of
the quaternionic non-linear $\sigma $-models we are interested in is
the $F_4$-model. This model (as well as all other ones) contains a
universal "hidden sector", based on the $O(4,4)$ group. The latter
has the $SU(2)^4$ as its maximal compact subgroup, whose generators
we will denote as $t_i{}^j$, $t_\alpha {}^\beta $,
$t_{\dot\alpha} {}^{\dot\beta} $ and $t_{\ddot\alpha} {}^{\ddot\beta}
$, correspondingly. Commutation relations for these generators are
normalized so that
\begin{equation}
 [ t_i{}^j, t_k{}^l ] = \delta _k{}^j t_i{}^l - \delta _i{}^l t_k{}^j
\end{equation}
and analogously for other ones. Besides, we will use the notation
$t_{ij} = \varepsilon _{jk} t_i{}^k = t_{ji}$ and so on. In this
basis the non-compact generators of the $O(4,4)$ group form a
multispinor $T_{i\alpha \dot\alpha \ddot\alpha }$, satisfying a
pseudo-reality condition
\begin{equation}
 (T_{i\alpha \dot\alpha \ddot\alpha })^* = T^{i\alpha \dot\alpha
 \ddot\alpha } = \varepsilon ^{ij}\varepsilon ^{\alpha \beta }
\varepsilon^ {\dot\alpha \dot\beta }\varepsilon^{\ddot\alpha
\ddot\beta } T_{j\beta \dot\beta \ddot\beta }.
\end{equation}
Commutation relations look like
\begin{eqnarray}
 [ t_i{}^j, T_{k\alpha \dot\alpha \ddot\alpha } ] &=& \delta _k{}^j
 T_{i\alpha \dot\alpha \ddot\alpha } - \frac12 \delta _i{}^j 
T_{k\alpha \dot\alpha \ddot\alpha } \quad {\rm plus\ similar\ ones\
for\ } \ \alpha , \dot\alpha , \ddot\alpha \nonumber \\
 \ [ T_{i\alpha \dot\alpha \ddot\alpha }, T^{j\beta \dot\beta
\ddot\beta } ]  &=& + \{ t_i{}^j \delta _\alpha{}^\beta \delta
_{\dot\alpha}{}^{\dot\beta}  \delta _{\ddot\alpha}{}^{\ddot\beta} +
\cdots \}.
\end{eqnarray}
Note that in our normalization the plus sign in the last commutator
corresponds to the noncompact group $O(4,4)$, while for the minus
sign one would have an $O(8)$ group.

   The key elements of the whole construction \cite{Ale75} are the
four commuting algebras of the form $[h,g] = 2g$. Let us choose
\begin{eqnarray}
 h_1 &=& (T_{1111} + T_{2222}), \qquad h_2 = (T_{1122} + T_{2211}),
\nonumber \\
 h_3 &=& (T_{1212} + T_{2121}), \qquad h_4 = (T_{1221} + T_{2112}).
\end{eqnarray}
Then one has
\begin{eqnarray}
 g_1 &=& \frac12 (t_{12}^{(1)} + t_{12}^{(2)} + t_{12}^{(3)} +
t_{12}^{(4)}) + \frac12 (T_{1111} - T_{2222}), \nonumber \\
 g_2 &=& \frac12 (t_{12}^{(1)} + t_{12}^{(2)} - t_{12}^{(3)} -
t_{12}^{(4)}) + \frac12 (T_{1122} - T_{2211}), \nonumber \\
 g_3 &=& \frac12 (t_{12}^{(1)} - t_{12}^{(2)} + t_{12}^{(3)} -
t_{12}^{(4)}) + \frac12 (T_{1212} - T_{2121}), \\
 g_4 &=& \frac12 (t_{12}^{(1)} - t_{12}^{(2)} - t_{12}^{(3)} +
t_{12}^{(4)}) + \frac12 (T_{1221} - T_{2112}), \nonumber
\end{eqnarray}
where $t^{(1)}$ stands for $t_{ij}$, $t^{(2)}$ for $t_{\alpha\beta}$
and so on. Besides the generators given above, algebra $O(4,4)$ (as
well as all algebras, corresponding to symmetric quaternionic spaces)
contains also four generators $\hat{g}$, such that $[h,\hat{g}] = -
2\hat{g}$, $[g,\hat{g}] = h$. They look like
\begin{eqnarray}
 \hat{g}_1 &=& \frac12 (t_{12}^{(1)} + t_{12}^{(2)} + t_{12}^{(3)} +
t_{12}^{(4)}) - \frac12 (T_{1111} - T_{2222}), \nonumber \\
 \hat{g}_2 &=& \frac12 (t_{12}^{(1)} + t_{12}^{(2)} - t_{12}^{(3)} -
t_{12}^{(4)}) - \frac12 (T_{1122} - T_{2211}), \nonumber \\
 \hat{g}_3 &=& \frac12 (t_{12}^{(1)} - t_{12}^{(2)} + t_{12}^{(3)} -
t_{12}^{(4)}) - \frac12 (T_{1212} - T_{2121}), \\
 \hat{g}_4 &=& \frac12 (t_{12}^{(1)} - t_{12}^{(2)} - t_{12}^{(3)} +
t_{12}^{(4)}) - \frac12 (T_{1221} - T_{2112}). \nonumber
\end{eqnarray}
Let us stress that it is the presence of all or some of the
generators $\hat{g}$ in the algebra that determines a possibility
to "restore" all or some of the four initial $SU(2)$ subgroups.

  Now one can combine all other generators of $O(4,4)$ algebra (as
well as all other generators of $F_4$) into the linear combinations
that will be the eihgenvectors for all four generators $h$. For
example, the 16 remaining generators of $O(4,4)$ form two octets with
$h_1$ eihgenvalues $\pm 1$:
\begin{eqnarray}
 \Omega_1^\pm &=& t_{22}^{(1)} \pm T_{2111} \qquad \Omega_2^\pm =
t_{22}^{(2)} \pm T_{1211}, \nonumber \\
 \Omega_3^\pm &=& t_{22}^{(3)} \pm T_{1121} \qquad \Omega_4^\pm =
t_{22}^{(4)} \pm T_{1112}, \nonumber \\
 \Omega_5^\pm &=& t_{11}^{(1)} \mp T_{1222} \qquad \Omega_6^\pm =
t_{11}^{(2)} \mp T_{2122},  \\
 \Omega_7^\pm &=& t_{11}^{(3)} \mp T_{2212} \qquad \Omega_8^\pm =
t_{11}^{(4)} \mp T_{2221}. \nonumber
\end{eqnarray}
 By combining them further into the linear combinations that are
eighenvectors for $h_2$, $h_3$ and $h_4$ one ends up with sixteen
combinations with the eighenvalues $\pm 1$ (see Appendix). Figure 1
shows the two-dimensional projection of this four-dimensional
diagram.
\begin{figure}[htb]
\setlength{\unitlength}{1mm}
\begin{picture}(150,120)
\put(10,60){\vector(1,0){130}}
\put(140,55){\makebox(10,10)[]{$h_1$}}
\multiput(15,60)(30,0){5}{\circle*{1}}
\put(10,60){\makebox(10,10)[]{$\hat{g}_1$}}
\put(60,60){\makebox(30,10)[]{$h_1,h_2,h_3,h_4$}}
\put(60,50){\makebox(30,10)[]{$\hat{g}_3,\hat{g}_4,g_3,g_4$}}
\put(130,60){\makebox(10,10)[]{$g_1$}}
\put(75,10){\vector(0,1){100}}
\put(70,110){\makebox(10,10)[]{$h_2$}}
\put(75,95){\makebox(10,10)[]{$g_2$}}
\put(75,15){\makebox(10,10)[]{$\hat{g}_2$}}
\multiput(75,20)(0,20){5}{\circle*{1}}
\put(75,60){\circle{2}}
\put(45,80){\circle*{1}}
\put(40,80){\makebox(10,10)[]{$\hat{T}_{+\pm\pm}$}}
\put(45,40){\circle*{1}}
\put(40,30){\makebox(10,10)[]{$\hat{T}_{-\pm\pm}$}}
\put(105,80){\circle*{1}}
\put(100,80){\makebox(10,10)[]{$T_{+\pm\pm}$}}
\put(105,40){\circle*{1}}
\put(100,30){\makebox(10,10)[]{$T_{-\pm\pm}$}}
\end{picture}
\caption{Generators of $O(4,4)$}.
\end{figure}
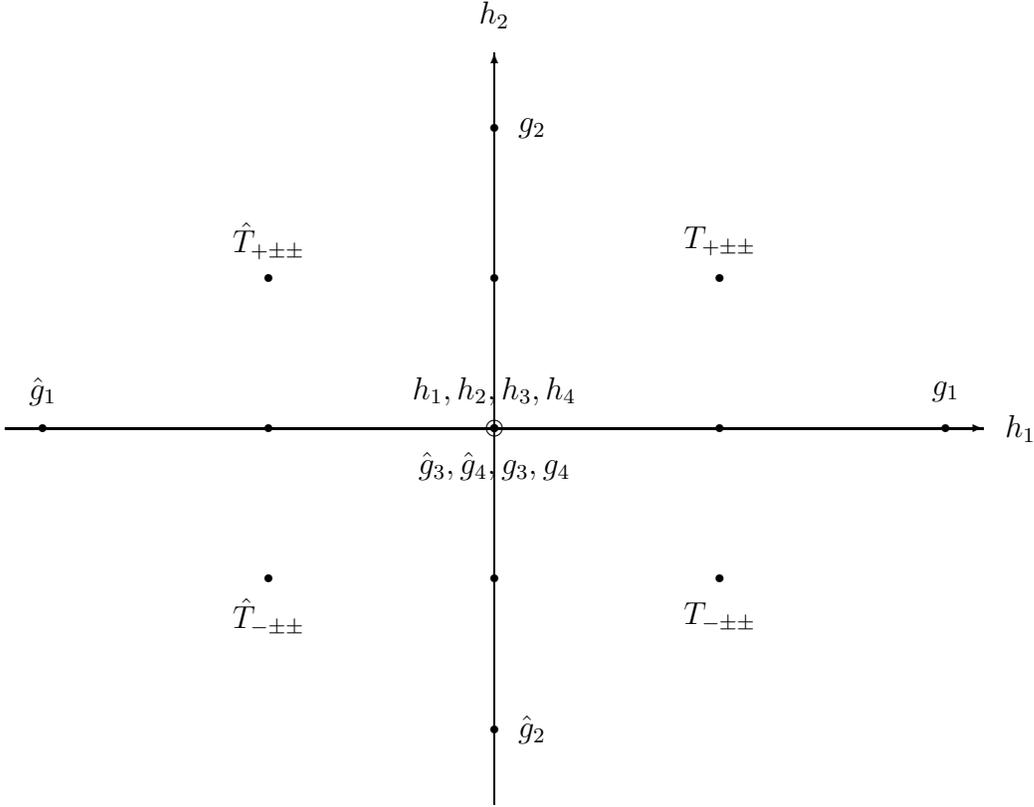
One can easily get convinced that sixteen generators $h$, $g$ and
$T_{\pm\pm\pm}$ form some closed algebra. It is this algebra that
determines the corresponding quaternionic manifold. Namely, for each
its generator one has scalar field in the appropriate nonlinear
$\sigma$-model playing the role of a Goldstone one. As for the
remaining generators all of them are present for the case of
symmetric quaternionic manifolds only. As we know \cite{Wit93,Wit94}
the generator $\hat{g}_1$ is present then and only then the manifold
is symmetric. But if we exclude this generator, one immediately see
that among eight generators $\hat{T}_{\pm\pm\pm}$ at most four
commuting generators $\hat{T}_{+\pm\pm}$ could present because their
commutators with $\hat{T}_{-\pm\pm}$ give $\hat{g}_1$. Moreover,
taking into account that $[ \hat{g}_2, \hat{T}_{+\pm\pm}] \sim
\hat{T}_{-\pm\pm}$ one must exclude the $\hat{g}_2$ as well. All this
means, in particular, that from the initial four $SU(2)$ subgroups
not more than two could survive in the cases of interest. Their
generators are formed by the linear combinations of $g_{3,4}$,
$\hat{g}_{3,4}$, $T_{+\pm\pm}$ and $\hat{T}_{+\pm\pm}$:
\begin{equation}
 t_a = \left( \begin{array}{c} t_{22}^1 + t_{11}^2 \\ t_{12}^1 -
t_{12}^2 \\ t_{11}^1 + t_{22}^2 \end{array} \right), \qquad
t_{\dot{a}} = \left( \begin{array}{c} t_{22}^3 + t_{11}^4 \\ t_{12}^3
- t_{12}^4 \\ t_{11}^3 + t_{22}^4 \end{array} \right),
\end{equation}
where $a,\dot{a} = 1,2,3$. The remaining six linear combinations
together with $h_{1,2,3,4}$ form a singlet $T = h_1 + h_2$ and
$(3,3)$ representation:
\begin{equation}
 T_{a\dot{a}} = \left( \begin{array}{ccc} T_{2121} & T_{2111} -
T_{2122} & T_{2112} \\ T_{1121} - T_{2221} & h_1 - h_2 & T_{1112} -
T_{2212} \\ T_{1221} & T_{1211} - T_{1222} & T_{1212} \end{array}
\right).
\end{equation}
In turn, the generators $g_{1,2}$ and $T_{+\pm\pm}$ form two triplets
$(3,1)$ and $(1,3)$:
\begin{equation}
 T_a = \left( \begin{array}{c} \Omega_1^+ - \Omega_6^+ \\ g_1 + g_2
\\ \Omega_2^+ - \Omega_5^+ \end{array} \right), \qquad T_{\dot{a}} =
\left(\begin{array}{c} \Omega_3^+ - \Omega_8^+ \\ g_1 - g_2 \\
\Omega_4^+ - \Omega_7^+ \end{array} \right).
\end{equation}
In this covariant under two $SU(2)$ groups notations the commutation
relations have the form:
\begin{eqnarray}
 [ T_a, T_b ] &=& 0, \qquad [ T_a, T_{\dot{b}} ] = 0, \qquad [
T_{\dot{a}},  T_{\dot{b}} ] = 0, \nonumber \\
 \  [ T, T_a ] &=& T_a, \qquad [ T, T_{\dot{a}} ] = T_{\dot{a}},
\qquad [ T, T_{a\dot{a}} ] = 0, \nonumber \\
 \  [ T_{a\dot{a}}, T_b ] &=& \delta_{ab} T_{\dot{a}}, \qquad [
T_{a\dot{a}}, T_{\dot{b}} ] = \delta_{\dot{a}\dot{b}} T_a, \\
 \ [ T_{a\dot{a}}, T_{b\dot{b}} ] &=& \varepsilon_{abc} t^c
\delta_{\dot{a}\dot{b}} + \delta_{ab}
\varepsilon_{\dot{a}\dot{b}\dot{c}} t^{\dot{c}}. \nonumber
\end{eqnarray}
From these relations we see that it is just the $O(3,3) \otimes D
\otimes T_{3,3}$, where $T_{3,3}$ --- six translations. Let us stress
that it is this subalgebra of the whole $O(4,4)$ algebra that
"survives" for all the quaternionic manifilds we are considering. For
what follows it will be useful to note that as a result of $O(3,3)
\simeq SL(4)$ this algebra is equivalent to $GL(4) \otimes T_6$,
where the generators $t_a$, $t_{\dot{a}}$, $T_{a\dot{a}}$ and $T$
form the $GL(4)$ algebra, while six translations are transformed as
skew-symmetric tensor $\Pi^{mn}$, $m,n=1,2,3,4$:
\begin{eqnarray}
  [ T_m{}^n, T_k{}^l ] &=& \delta_m{}^l T_k{}^n - \delta_k{}^n
T_m{}^l, \nonumber \\
 \  [ T_m{}^n, \Pi^{kl} ] &=& \delta_m{}^k \Pi^{nl} - \delta_m{}^l
\Pi^{nk}, \\
 \  [ \Pi^{mn}, \Pi^{kl} ] &=& 0. \nonumber
\end{eqnarray}

 Now let us turn to the whole group $F_4$. The noncompact version of
this group that we need contains as a maximal regular (noncompact)
subgroup an $O(4,5)$ group formed by the $O(4,4)$ group described
above and by the eight generators
$(\Lambda_{i\alpha},\Lambda_{\dot{\alpha}\ddot{\alpha}})$. The
remaining sixteen generators are transformed as a spinor
representation of this $O(4,5)$ group and in our basis take the form:
$(\Lambda_{i\dot{\alpha}}, \Lambda_{i\ddot{\alpha}},
\Lambda_{\alpha\dot{\alpha}}, \Lambda_{\alpha\ddot{\alpha}})$. Note,
that all these $\Lambda$-generators are complex ones, satisfying the
pseudoreality condition $(\Lambda_{i\alpha})^* = \Lambda^{i\alpha} =
\varepsilon^{ij} \varepsilon^{\alpha\beta} \Lambda_{j\beta}$ and so
on.  All commutation relations for these generators are given in
Appendix. In the same way as for the $O(4,4)$ generators we can
construct linear combinations which are eighenvectors for the
$h_{1,2,3,4}$. The explicit form of such combinations also given in
the Appendix. From 24 generators we get 12 combinations with $h_1 =
0$ and 6 combinations with $h_1 = 1$ (see Figure 2) as well as 6
combinations with $h_1 = -1$ which we denote as
$\hat{\tilde{X}}_{\pm}$, $\hat{\tilde{Y}}_{\pm}$ and
$\hat{\tilde{Z}}_{\pm}$.
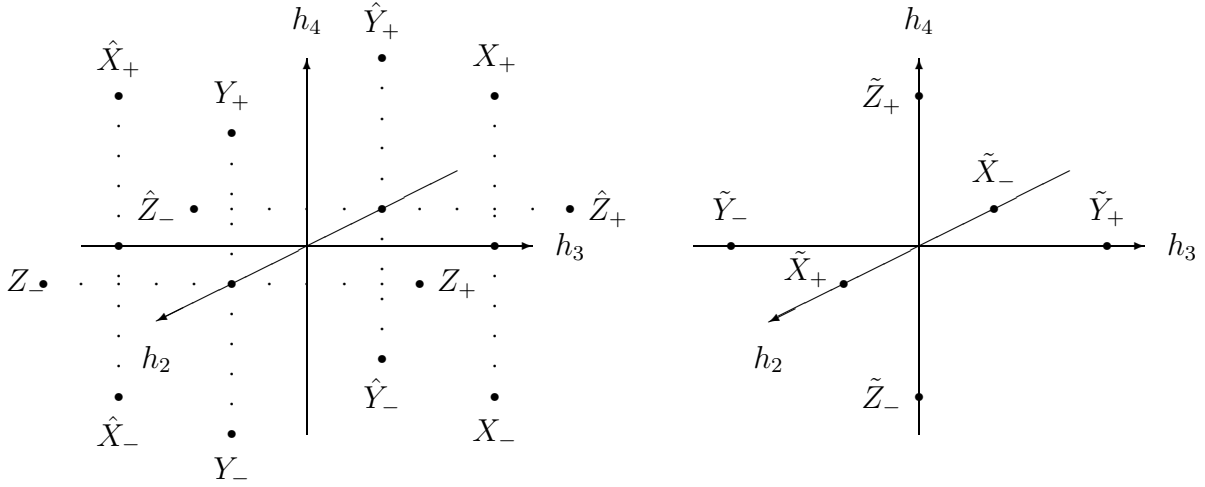
\begin{figure}[htb]
\setlength{\unitlength}{1mm}
\noindent
\begin{picture}(80,80)
\put(10,40){\vector(1,0){60}}
\put(70,35){\makebox(10,10)[]{$h_3$}}
\put(15,40){\circle*{1}}
\put(65,40){\circle*{1}}
\put(10,60){\makebox(10,10)[]{$\hat{X}_+$}}
\put(15,60){\circle*{1}}
\multiput(15,24)(0,4){9}{\circle*{0.5}}
\put(15,20){\circle*{1}}
\put(10,10){\makebox(10,10)[]{$\hat{X}_-$}}
\put(60,60){\makebox(10,10)[]{$X_+$}}
\put(65,60){\circle*{1}}
\multiput(65,24)(0,4){9}{\circle*{0.5}}
\put(65,20){\circle*{1}}
\put(60,10){\makebox(10,10)[]{$X_-$}}

\put(40,15){\vector(0,1){50}}
\put(35,65){\makebox(10,10)[]{$h_4$}}

\put(60,50){\vector(-2,-1){40}}
\put(15,20){\makebox(10,10)[]{$h_2$}}
\put(50,45){\circle*{1}}
\put(30,35){\circle*{1}}

\put(45,65){\makebox(10,10)[]{$\hat{Y}_+$}}
\put(50,65){\circle*{1}}
\multiput(50,29)(0,4){9}{\circle*{0.5}}
\put(50,25){\circle*{1}}
\put(45,15){\makebox(10,10)[]{$\hat{Y}_-$}}
\put(15,40){\makebox(10,10)[]{$\hat{Z}_-$}}
\put(25,45){\circle*{1}}
\multiput(30,45)(5,0){9}{\circle*{0.5}}
\put(75,45){\circle*{1}}
\put(75,40){\makebox(10,10)[]{$\hat{Z}_+$}}

\put(0,30){\makebox(10,10)[l]{$Z_-$}}
\put(5,35){\circle*{1}}
\multiput(10,35)(5,0){9}{\circle*{0.5}}
\put(55,35){\circle*{1}}
\put(55,30){\makebox(10,10)[]{$Z_+$}}
\put(25,5){\makebox(10,10)[]{$Y_-$}}
\put(30,15){\circle*{1}}
\multiput(30,19)(0,4){9}{\circle*{0.5}}
\put(30,55){\circle*{1}}
\put(25,55){\makebox(10,10)[]{$Y_+$}}
\end{picture}
\begin{picture}(80,80)
\put(10,40){\vector(1,0){60}}
\put(70,35){\makebox(10,10)[]{$h_3$}}
\put(15,40){\circle*{1}}
\put(65,40){\circle*{1}}
\put(10,40){\makebox(10,10)[]{$\tilde{Y}_-$}}
\put(60,40){\makebox(10,10)[]{$\tilde{Y}_+$}}

\put(40,15){\vector(0,1){50}}
\put(35,65){\makebox(10,10)[]{$h_4$}}
\put(40,20){\circle*{1}}
\put(40,60){\circle*{1}}
\put(30,15){\makebox(10,10)[]{$\tilde{Z}_-$}}
\put(30,55){\makebox(10,10)[]{$\tilde{Z}_+$}}

\put(60,50){\vector(-2,-1){40}}
\put(15,20){\makebox(10,10)[]{$h_2$}}
\put(50,45){\circle*{1}}
\put(30,35){\circle*{1}}
\put(45,45){\makebox(10,10)[]{$\tilde{X}_-$}}
\put(20,32){\makebox(10,10)[]{$\tilde{X}_+$}}

\end{picture}
\caption{Generators of $F_4/O(4,4)$ with $h_1=0$ (left) and $h_1=1$
(right)}.
\end{figure}
Note, the letters $X,Y,Z$ are chosen so that they match the
Alekseevsky notations in \cite{Ale75}. As we have seen, the $F_4$
algebra has just one set of each type of generators (that is why we
choose it as our starting point). In this, together with the
generators $h_{1,2,3,4}$, $g_{1,2,3,4}$ and $T_{\pm\pm\pm}$,
described before, the generators $\{X_\pm, \tilde{X}_\pm, Y_\pm,
\tilde{Y}_\pm, Z_\pm, \tilde{Z}_\pm \}$ form the closed algebra with
dimension 28, corresponding to this quaternionic manifold. 

  In general, the quaternionic algebra can contain different numbers
of $X,Y,Z$ generators, the whole dimensions of the algebras being, of
course, multiples of four. As it has been shown \cite{Ale75,Cec89}
there exist two general types of quaternionic manifolds which are not
symmetric ones. In the first one the $X$ type generators are absent,
while one can have the arbitrary number $p$ of the sets $(Y_{\pm},
\tilde{Y}_{\pm})$ and the arbitrary number $q$ of the $(Z_{\pm},
\tilde{Z}_{\pm})$. In the second case we have $q$ sets of $X$-type
generators, while the quaternionic dimensions of $Y$'s and $Z$'s are
both equal to the $p \times d(q)$, where $p$ is arbitrary and $d(q)$
is the dimension of the spinor representation of $O(q)$. In the
following sections we will give explicit realizations for such
algebras and construct the corresponding $N=2$ supersymmetric
nonlinear $\sigma$-models.

\section{$W(p,q)$--model}

  Let us first consider the quaternionic algebras which arise in the
absence of the $X$-type generators. Under the two "surviving" $SU(2)$
subgroups the generators $(Y_{\pm}, \tilde{Y}_{\pm})$ and $(Z_{\pm},
\tilde{Z}_{\pm})$, defined previously are transformed as the
bispinors $Y_{i\alpha}$ and $Z_{i\alpha}$ (note that in this section
indices $i,\alpha$ correspond to the two new $SU(2)$ groups and do
not coincide with the ones used before). The commutation relations
look like:
\begin{eqnarray}
[ T_{a\dot{a}}, Y_{i\alpha} ] &=& (\sigma^a)_i{}^j
 (\sigma^{\dot{a}})_{\alpha}{}^{\beta}Y_{j\beta}, \qquad
 [ T_{a\dot{a}}, Z_{i\alpha} ] = - (\sigma^a)_i{}^j
(\sigma^{\dot{a}})_{\alpha}{}^{\beta} Z_{j\beta}, \nonumber \\
 \ [ Y_{i\alpha}, Y_{j\beta} ] &=& (\sigma^a)_{ij}
\varepsilon_{\alpha\beta} T_a + \varepsilon_{ij}
(\sigma^{\dot{a}})_{\alpha\beta} T_{\dot{a}}, \nonumber \\
 \ [ Z_{i\alpha}, Z_{j\beta} ] &=& - (\sigma^a)_{ij}
\varepsilon_{\alpha\beta} T_a + \varepsilon_{ij}
(\sigma^{\dot{a}})_{\alpha\beta} T_{\dot{a}}, \\
  \ [ T_a, Y_{i\alpha} ] &=& [ T_{\dot{a}}, Y_{i\alpha} ] = [ T_a,
Z_{i\alpha} ] = [ T_{\dot{a}}, Z_{i\alpha} ] = 0, \nonumber \\
 \ [ T, Y_{i\alpha} ] &=& Y_{i\alpha} \qquad [ T, Z_{i\alpha} ] =
Z_{i\alpha}, \nonumber
\end{eqnarray}
where $\sigma^a$, $\sigma^{\dot{a}}$ are Pauli matrices.
Under the $GL(4)$ group mentioned at the end of the previous section
we have covariant and contravariant vectors $Y^m$ and $Z_m$,
correspondingly. Then the algebra takes a very simple form:
\begin{eqnarray}
 [ T_m{}^n, Y^k ] &=& - \delta_m{}^k Y^n, \qquad [ T_m{}^n, Z_k ] =
\delta_k{}^n Z_m, \nonumber \\
 \ [ Y^m, Y^n ] &=& \Pi^{mn}, \qquad [ Z_m, Z_n ] = \frac12
\varepsilon_{mnkl} \Pi^{kl}, \\
 \ [ \Pi^{mn}, Y^k ] &=& 0, \qquad [ \Pi^{mn}, Z_k ] = 0, \qquad [
Y^m, Z_n ] = 0. \nonumber
\end{eqnarray}
Now, it is easy to see that there is nothing to prevent us from
considering an arbitrary number of generators $Y^{n\dot{A}}$,
$\dot{A}=1,2\cdots p$ and $Z_N{}^{\ddot{A}}$, $\ddot{A}=1,2\cdots q$.
The commutation relations remain the same except that one now can
introduce new generators $t^{[\dot{A}\dot{B}]}$ and
$t^{[\ddot{A}\ddot{B}]}$, corresponding to $O(p)$ and $O(q)$ groups
with evident commutation relations. In this, the generators that we
have excluded are transformed as $\hat{Y}_n$ and $\hat{Z}^n$,
correspondingly. It appears that in the case when, say, $q=0$, i.e.
the generators of $Z$-type are absent, we may include the generators
$\hat{Y}_n$ into the algebra (together with the previously excluded
$\hat{g}_{1,2}$ and $T_{-\pm\pm}$) extending the algebra up to the
full $O(4,4+p)$ one. Thus, in the absence of $Z$-type generators we
have just the well known $O(4,4+p)/O(4)\otimes O(4+p)$ nonlinear
$\sigma$-model. It allows one to get a very simple description of
such a model which turns out to be convenient for the generalization
on the $q>0$ case.

   The most simple way to describe the required $\sigma$-model is to
introduce the $p+8$ hypermultiplets $(\Phi_a{}^{\hat{A}},
\Lambda_{\alpha}{}^{\hat{A}})$, $a=1,2,3,4$, $\hat{A}=1,2\cdots p+8$,
$\alpha=1,2$, where $\Phi$ are the real scalar fields and $\Lambda$
--- Majorana spinors, satisfying the following constraints:
\begin{equation}
 \Phi_a{}^{\hat{A}} \cdot \Phi_{b\hat{A}} = - \delta_{ab}, \qquad
\Phi_a{}^{\hat{A}} \cdot \Lambda_{\alpha\dot{A}} = 0.
\end{equation}
In this, the theory has the local $O(4)$ invariance, the
corresponding covariant derivatives being, for example:
\begin{eqnarray}
 D_\mu \Phi_a{}^{\hat{A}} &=& \partial_\mu \Phi_a{}^{\hat{A}} -
(\Phi_a \partial_\mu \Phi_b) \Phi_b{}^{\hat{A}}, \qquad \Phi_a D_\mu
\Phi_b = 0, \nonumber \\
 D_\mu \Lambda^{\hat{A}} &=& \partial_\mu \Lambda^{\hat{A}} + \frac12
(\Phi_a \partial_\mu \Phi_b) \bar{\sigma}^{ab} \Lambda^{\hat{A}}.
\end{eqnarray}
In this notations the Lagrangian of the interaction of such
hypermultiplets with $N=2$ supergravity has the form:
\begin{eqnarray}
 {\cal L} &=& {\cal L}_{N=2 sugra} + {\cal L}_{hyper}, \nonumber \\
 {\cal L}_{hyper} &=& \frac{i}2 \bar{\Lambda} \gamma^\mu D_\mu
\Lambda + \frac12 D_\mu \Phi_a D^\mu \Phi_a - \frac12 \bar{\Lambda}
\gamma^\mu \gamma^\nu D_\nu \Phi_a \bar{\tau}^a \Psi_\mu.
\end{eqnarray}
Here we introduced four matrices $(\tau_a)_{i\alpha}$ and
$(\bar{\tau}_a)^{\alpha i}$ such that:
\begin{equation}
 (\tau^a)_{i\alpha} (\bar{\tau}^b)^{\alpha j} + (a \leftrightarrow b)
= 2 \delta^{ab} \delta_i{}^j
\end{equation}
in this,
\begin{equation}
 (\sigma^{ab})_i{}^j = \frac12 ((\tau^a)_{i\alpha}
(\bar{\tau}^b)^{\alpha j} - (a \leftrightarrow b)), \qquad
  (\bar{\sigma}^{ab})_{\alpha}{}^{\beta} = \frac12
((\tau^a)_{i\alpha} (\bar{\tau}^b)^{\beta i} - (a \leftrightarrow
b)).
\end{equation}

   Now let us rewrite this Lagrangian and the supertransformations in
terms of independent scalar and spinor fields. For that purpose we
introduce a kind of light cone variables:
\begin{equation}
 \Phi_a{}^{\hat{A}} = (x_a{}^m + E_{am}, x_a{}^m - E_{am}, E_{am}
Y^{m\dot{A}}), \quad m = 1,2,3,4, \quad \dot{A} = 1,2 \cdots p.
\end{equation}
Now, by introducing a new field
\begin{equation}
 \Pi^{mn} = (E^{-1})^{am} x_a{}^n - (E^{-1})^{an} x_a{}^m
\end{equation}
we can solve the constraint for $x_a{}^m$:
\begin{equation}
 x_a{}^m = \frac12 (E^{-1})^{am} + \frac14 E_{an} Y^{n\dot{A}}
Y^{m\dot{A}} + \frac12 E_{an} \Pi^{nm}
\end{equation}
and rewrite all the bosonic expressions in terms of the $E_{am}$,
$\Pi^{mn}$ and $Y^{m\dot{A}}$. In this, the fields $E_{am}$ realise
the nonlinear $\sigma$-model $GL(4)/O(4)$, while $\Pi^{mn}$ enter
the Lagrangian through the derivatives $\partial_\mu \Pi^{mn}$ only,
the translations $\Pi^{mn} \rightarrow \Pi^{mn} + \Lambda^{mn}$ being
the global symmetry.

  Analogously, we solve the constraint for the spinor fields
introducing new fields $\Lambda^{\hat{A}} = (\xi^m + \chi_m, \xi^m -
\chi_m, \Lambda^{\dot{A}})$. This gives
\begin{equation}
 \xi^m = - (E^{-1})^{am} x_a{}^n \chi_n + \frac12 Y^{m\dot{A}}
\Lambda_{\dot{A}}.
\end{equation}
Here two changes of variables $\chi_m \rightarrow \frac1{\sqrt{2}}
E_{am} \chi^a$ and $\Lambda \rightarrow \Lambda + Y^m \chi_m$ are
necessary to have canonical kinetic terms for spinors.

  In terms of these new variables the Lagrangian could be written as:
\begin{eqnarray}
 {\cal L}_{hyper} &=& \frac{i}2 \bar{\chi} \gamma^\mu D_\mu \chi +
\frac12 (S_\mu^+)^2 + \frac12 (P_\mu)^2 + \frac{i}2 \bar{\Lambda}
\gamma^\mu D_\mu \Lambda + \frac12 E_{ma} E_{na}
\partial_\mu Y^m \partial_\mu Y^n \nonumber \\
 && - \frac12 \bar{\chi}^a \gamma^\mu \gamma^\nu (S_\nu{}^+ +
P_\nu)_{ab} \bar{\tau}^b \Psi_\mu - \frac12 \bar{\Lambda} \gamma^\mu
\gamma^\nu E_{ma} \partial_\nu Y^m \bar{\tau}^a \Psi_\mu \nonumber \\
 && - \frac{i}8 \varepsilon^{\mu\nu\rho\sigma} \bar{\Psi}_\mu
\gamma_5 \gamma_\nu (S_\rho{}^- + P_\rho)_{ab} \sigma^{ab}
\Psi_\sigma + \frac{i}8 \bar{\chi}^c \gamma^\mu (S_\mu^- +
P_\mu)_{ab} \bar{\sigma}^{ab} \chi^c \nonumber \\
 && - \frac{i}2 \bar{\chi}^a \gamma^\mu (S_\mu^- - P_\mu)_{ab} \chi^b
+ \frac{i}8 \bar{\Lambda} \gamma^\mu (S_\mu^- - P_\mu)_{ab}
\bar{\sigma}^{ab} \Lambda + \frac{i}2 \bar{\chi}^a \gamma^\mu E_{ma}
\partial_\mu Y^m \Lambda.
\end{eqnarray}
Here we use the following notations:
\begin{equation}
 (S_\mu^\pm)_{ab} = \frac12 (E^{-1} \partial_\mu E \pm \partial_\mu E
E^{-1})_{ab}, \qquad (P_\mu)_{ab} = E_{ma} E_{nb} ( \partial_\mu
\Pi^{mn} - \frac12 (Y^m \stackrel{\leftrightarrow}{\partial}_\mu
Y^n)).
\end{equation}
In turn, the supertransformations look like:
\begin{eqnarray}
 \delta \Psi_\mu &=& 2 D_\mu \eta - \frac12 (S_\mu^- + P_\mu)_{ab}
\sigma^{ab} \eta, \nonumber \\
 \delta \chi^a &=& - i \gamma^\mu (S_\mu^+ + P_\mu)_{ab} \bar{\tau}^b
\eta, \qquad \delta \Lambda = - i \gamma^\mu E_{ma} \partial_\mu Y^m
\bar{\tau}^a \eta, \nonumber \\
 \delta E_{ma} &=& (\bar{\chi}^b E_{bm} \bar{\tau}^a \eta), \qquad
\qquad \quad \delta Y^m = (\bar{\Lambda} (E^{-1})^{ma} \bar{\tau}^a
\eta), \\
 \delta \Pi^{mn} &=& (\bar{\chi}^a (E^{-1})_a{}^{[m}
(E^{-1})_b{}^{n]} \bar{\tau}^b \eta) + (\bar{\Lambda} Y^{[m}
(E^{-1})^{n]a} \bar{\tau}^a \eta). \nonumber
\end{eqnarray}
This Lagrangian, besides the $GL(4)$ transformations (acting on the
"world" indices $m,n$ only), is invariant under the six translations
$\Pi^{mn} \rightarrow \Pi^{mn} + \Lambda^{mn}$ as well as under the
following transformations:
\begin{equation}
 \delta Y^m = \xi^m, \qquad \delta \Pi^{mn} = Y^{[m} \xi^{n]}.
\end{equation}

  Now it is an easy task to add to this model additional
hypermultiplets whose scalar fields are transformed as
$Z_m{}^{\ddot{A}}$, $\ddot{A} = 1,2 \cdots q$, under $GL(4)$. All
one needs for that is to complete the Lagrangian with:
\begin{eqnarray}
 \Delta L_{hyper} &=& \frac{i}2 \bar{\Sigma} \gamma^\mu D_\mu \Sigma
+  \frac{\Delta}2 (E^{-1})^{am} (E^{-1})^{an} \partial_\mu Z_m
\partial_\mu Z_n \nonumber \\
 && - \frac12 \bar{\Sigma} \gamma^\mu \gamma^\nu \sqrt{\Delta}
(E^{-1})^{am} \partial_\nu Z_m \bar{\tau}^a \Psi_\mu + \frac{i}8
\bar{\Sigma} \gamma^\mu (S_\mu^- - P_\mu)_{ab} \bar{\sigma}^{ab}
\Sigma \nonumber \\
 && + \frac{i}2 \bar{\chi}^a \gamma^\mu \sqrt{\Delta} (E^{-1})^{am}
\partial_\mu Z_m \Sigma + \frac{i}2 \bar{\chi}^a \gamma^\mu
\sqrt{\Delta} (E^{-1})^{bm} \partial_\mu Z_n \bar{\sigma}^{ab}
\Sigma,
\end{eqnarray}
where $\Delta = det(E_{ma})$, and the supertransformations with
\begin{eqnarray}
 \delta \Sigma &=& - i \gamma^\mu \sqrt{\Delta} (E^{-1})^{am}
\partial_\mu Z_m \bar{\tau}^a \eta, \nonumber \\
 \delta Z_m &=& (\bar{\Sigma}
\sqrt{\Delta} (E^{-1})^{am} \bar{\tau}^a \eta), \\
 \delta' \Pi^{mn} &=& \frac1{2\sqrt{\Delta}} (\bar{\Sigma}
\varepsilon^{mnpq} E_{pa} Z_q \bar{\tau}^a \eta). \nonumber
\end{eqnarray}
Besides, one has to change the definition for the $P_\mu$ by
\begin{equation}
  (P_\mu)_{ab} = E_{am} E_{bn} (\partial_\mu \Pi^{mn} - \frac12 (Y^m
\stackrel{\leftrightarrow}{\partial_\mu} Y^n) + \frac14
\varepsilon^{mnpq} (Z_p \partial_\mu Z_q)).
\end{equation}
Now, besides the bosonic transformations given above, the whole
Lagrangian is also invariant under:
\begin{equation}
 \delta Z_m = \eta_m, \qquad \delta \Pi^{mn} = \frac12
\varepsilon^{mnpq} Z_p \eta_q.
\end{equation}
It is an easy task to check that these bosonic transformations have
the commutation relations which coincide with the ones given at the
beginnig of this section. Thus the bosonic part of the Lagrangian
constructed is indeed the nonlinear $\sigma$-model, corresponding to
the quaternionic manifold of the desired form.

\section{$V(p,0)$--model}

  As a preliminary step to the full $V(p,q)$--model, let us consider
first one more simple case --- the one without $Y$ and $Z$
multiplets. Under the two "surviving" $SU(2)$ subgroups the $X$
generators form two triplets $X_a$, $X_{\dot{a}}$ and two singlets
$X$ and $\tilde{X}$. In this, the minimal quaternionic algebra could
be extended so that to include the $X_a$, $X_{\dot{a}}$ and $X$
generators together with the $O(3,3)\otimes D \otimes T_{3,3}$ ones,
defined earlier. The commutation relations have the form:
\begin{eqnarray}
  [ T_{a\dot{a}}, X_b ] &=& \delta_{ab} X_{\dot{a}}, \qquad
 [ T_{a\dot{a}}, X_{\dot{b}} ] = \delta_{\dot{a}\dot{b}} X_a,
\nonumber \\
 \ [ X_a, X_b ] &=& - \varepsilon_{abc} t^c \qquad [ X_{\dot{a}},
X_{\dot{b}} ] = \varepsilon_{\dot{a}\dot{b}\dot{c}} t^{\dot{c}},
\qquad [ X_a, X_{\dot{a}} ] = - T_{a\dot{a}}, \\
 \ [ X_a, T_b ] &=& \delta_{ab} X, \qquad [ X_{\dot{a}}, T_{\dot{b}}
] = - \delta_{\dot{a}\dot{b}} X, \qquad  [ X_a, X ] = T_a, \qquad
 [ X_{\dot{a}}, X ] = T_{\dot{a}}. \nonumber
\end{eqnarray}
From these relations one can see that the generators $X_a$ and
$X_{\dot{a}}$ together with the $t_a$, $t_{\dot{a}}$ and
$T_{a\dot{a}}$ form the $O(3,4)$ algebra, while the $X$ generators
together with the $T_a$ and $T_{\dot{a}}$ form seven translations
$T_{3,4}$. Moreover, there is nothing to prevent us from considering
the generalization on the case of arbitrary number $p$ of generators
$X_a$, $X_{\dot{a}}$ and $X$. In this case the algebra consists of
the $O(3,p+3)$ group, scale transformations $D$, and $p+6$
translations. In the absence of the $Y$, $Z$ multiplets this algebra
can be completed up to the whole $O(4,p+4)$ one. So, we start once
more with the usual non-linear $\sigma$-model $O(4,p+4)/O(4)\otimes
O(p+4)$ described by the real scalar fields
$\Phi_{\hat{a}}{}^{\hat{A}}$, $\hat{a}=1,2,3,4$, $\hat{A} = 0,1,2
\cdots p+7$, but this time (in order to preserve linearly realized
$O(3,3+p)$ symmetry) we will solve the constraint
$\Phi_{\hat{a}} \Phi_{\hat{b}} = - \delta_{\hat{a}\hat{b}}$ only
partially introducing the following parameterization: 
\begin{equation}
 \Phi_{\hat{a}}{}^{\hat{A}} = \left( \begin{array}{c|c|c} \phi_- &
X^A & \phi_+ \\ \hline L_a^- & L_a{}^A & L_a^+ \end{array} \right),
\end{equation}
where now $a=1,2,3$ and $A=1,2, \cdots p+6$. In this notations the
constraint takes the form:
\begin{eqnarray}
 - L_a^- L_b^- + (L_a L_b) + L_a^+ L_b^+ &=& - \delta_{ab}, \nonumber
\\
 - (\phi_-)^2 + (X)^2 + (\phi_+)^2 &=& - 1, \\
 - \phi_- L_a^- + (L_a X) + \phi_+ L_a^+ &=& 0. \nonumber
\end{eqnarray}
As usual, we have an $O(4) \simeq O(3) \otimes O(3)$ local
invariance, so one can use one of these $O(3)$ groups in order to set
$L_a^+ = L_a^-$. Then the first equation becomes $(L_a L_b) = -
\delta_{ab}$, i.e. just the usual constraint for the $O(3,p+3)/O(3)
\otimes O(p+3)$ non-linear $\sigma$-model! Two other equations allow
one to rewrite all the formulas in terms of $L_a{}^A$, $X^A$ and,
say, $\Phi = (\phi_- - \phi_+)$. The bosonic part of the
corresponding quaternionic model looks like:
\begin{equation}
 {\cal L}_B = \frac12 \frac{(\partial_\mu \Phi)^2}{\Phi^2} + \frac12
\Phi^2 ((\partial_\mu X)^2 + 2 (\vec{L} \partial_\mu X)^2) + \frac12
D_\mu \vec{L} D_\mu \vec{L},
\end{equation}
where $D_\mu$ is $O(3)$ covariant derivative. As we see the $X$
scalar fields enter the Lagrangian through the derivatives only, so
the Lagrangian is trivially invariant under the translations $X^A
\rightarrow X^A + \Lambda^A$, as well as $O(3,p+3)$ rotations and
scale transformations. By rather long but strightforward calculations
one can check that this Lagrangian is invariant also under the
"special conformal" transformations of the form:
\begin{eqnarray}
 \delta \Phi &=& \Phi (X \Lambda), \qquad \delta \vec{L}^A = (\vec{L}
X) \Lambda^A - (\vec{L} \Lambda) X^A, \nonumber \\
 \delta X^A &=& \frac12 \Phi^{-2} \Lambda^A - (X \Lambda) X^A +
\frac12 (X X) \Lambda^A + \Phi^{-2} (\vec{L} \Lambda) \vec{L}^A,
\end{eqnarray}
which, together with the ones mentioned above, form the full
$O(4,p+4)$ algebra. 

  Analogously, by solving partially the constraint for the spinor
fields and making field redefinitions to bring the fermionic kinetic
terms to the canonical forms, one can express the fermionic part of
the corresponding quaternionic model in terms of the spinors $\chi^i$
and $\Omega^{iA}$, satisfying $\vec{L}^A \Omega^{iA} = 0$. The
results are:
\begin{eqnarray}
 {\cal L}_F &=& - \frac12 \bar{\Omega}^i \gamma^\mu \gamma^\nu [ \Phi
(\partial_\nu X + \vec{L} (\vec{L} \partial_\nu X)) \delta_i{}^j +
D_\nu L_i{}^j ] \Psi_{\mu j} \nonumber \\
 && - \frac12 \bar{\chi}^i \gamma^\mu \gamma^\nu (\Phi^{-1}
\partial_\nu \Phi \delta_i{}^j - \Phi (L_i{}^j \partial_\nu X))
\Psi_{\mu j} + \frac{i}4 \Phi \varepsilon^{\mu\nu\rho\sigma}
\bar{\Psi}_\mu{}^i \gamma_5 \gamma_\nu (L_i{}^j \partial_\rho X)
 \Psi_{\sigma j} \nonumber \\
 && +\frac{i}4 \Phi \bar{\chi}_i \gamma^\mu (L_i{}^j \partial_\mu X)
 \chi^j + \frac{i}4 \Phi \bar{\Omega}_i \gamma^\mu (L_i{}^j
\partial_\mu X) \Omega^j - i \Phi \bar{\chi}_i \gamma^\mu
\partial_\mu X \Omega^i
\end{eqnarray}
and
\begin{eqnarray}
 \delta \Psi_{\mu i} &=& 2 D_\mu \eta_i + \Phi (L_i{}^j \partial_\mu
X) \eta_j, \nonumber \\
 \delta \chi^i &=& - i \gamma^\mu [ \Phi^{-1} \partial_\mu \Phi
\delta_i{}^j - \Phi (L_i{}^j \partial_\mu X)] \eta_j, \nonumber \\
 \delta \Omega^i &=& - i \gamma^\mu [ \Phi (\partial_\mu X + \vec{L}
(\vec{L} \partial_\mu X)) \delta_i{}^j + D_\mu L_i{}^j ] \eta_j, \\
 \delta \Phi &=& \Phi (\bar{\chi}^i \eta_i) \qquad \quad \delta
\vec{L} = (\bar{\Omega}^i (\vec{\tau})_i{}^j \eta_j), \nonumber \\
 \delta X &=& \Phi^{-1} [(\bar{\Omega}^i \eta_i) + (\bar{\chi}^i
L_i{}^j \eta_j)]. \nonumber
\end{eqnarray}

\section{$V(p,q)$--model}

   Thus, both for the $YZ$-sector and for the $X$-sector we have
managed to construct rather simple realizations having essentially
the same hidden sector (the $O(4,4)/O(4)\otimes O(4)$ model) but in
the very different parameterizations. So, to join these models
together we have to reduce them to the forms having identical
parameterization. Let us start with the $X$-sector. In this case all
that we need is to solve the remaining constraints $L^a L^b = -
\delta^{ab}$ and $\vec{L} \Omega = 0$ exactly in the same way as we
have done it for the initial $O(4,m)$ model above. In this, the
fields $\vec{L}^A$ give $y_{ma}$, $\pi^{[mn]}$ and $X^{mA}$, where
now $A = 7,8,\dots p+3$, $a,m = 1,2,3$, while $X^A$ give $\l^m$,
$\pi_m$ and $X^A$, correspondingly. The bosonic Lagrangian looks
like
\begin{eqnarray}
{\cal L}^B &=& \frac12 (\partial_{\mu}\varphi)^2 + \frac12 (S_{\mu
ab}^+)^2 + \frac12 (P_{\mu ab})^2 + 4e^{2\varphi} g_{mn} L_{\mu}^n
L_{\nu}^n + \nonumber \\
 && + \frac14 e^{2\varphi} g^{mn} U_{\mu m} U_{\mu n} + \frac12
e^{2\varphi} (D_{\mu}X)^2+\frac{1}{2}g_{mn}
\partial_{\mu}X^m\partial_{\mu}X^n, \label{e1}
\end{eqnarray}
where $g_{mn}=y_{ma}y_{na}$, $g^{mn}=y^{ma}y^{na}$, $\Phi =
e^\varphi$ and we have introduced the following notations:
\begin{eqnarray}
 S^{\pm}_{\mu ab} &=& \frac12 [y^{ma} \partial_{\mu} y_{mb} \pm
y^{mb} \partial_{\mu} y_{ma}], \qquad U_{\mu m}=\partial_{\mu}\pi_m,
\nonumber \\
 L^m_{\mu} &=& \partial_{\mu} \l^m + \frac12 \pi^{mn} U_{\mu n} +
\frac12 X^n D_{\mu} X - \frac14 X^m X^n U_{\mu n},  \nonumber   \\
 P_{\mu ab} &=& y_{ma} \left\{ \partial_{\mu} \pi^{mn} + \frac12
(X^m \stackrel{\leftrightarrow}{\partial_{\mu}} X^n) \right\} y_{nb},
\label{e2} \\
 Q^{a\pm}_{\mu} &=& y_{ma} L^m_{\mu} \pm \frac14 y^{ma} U_{\mu m},
\qquad D_{\mu} X = \partial_{\mu} X + X^m U_{\mu m}. \nonumber
\end{eqnarray}

  This Lagrangian, besides the $GL(3)$ group acting on the "world"
indices $m,n$, scale transformations and trivial translations for the
fields $\l^m$, $\pi_m$ and $X^A$, is invariant under the following
global transformations:
\begin{eqnarray}
 \delta X^m &=& \zeta^m, \quad \delta X = - \zeta^m \pi_m, \quad
\delta \pi^{mn} = X^{[m} \zeta^{n]}, \quad \delta \l^m = \frac12
\zeta^m X,  \nonumber  \\
 \delta \pi^{mn} &=& -2 \Lambda^{mn}, \qquad \delta \l^m =
\Lambda^{mn} \pi_n. \label{e3}
\end{eqnarray}

  Analogously, by solving the constraint for the spinor field in
terms of $(\lambda^{ia}, \Omega^{iA})$ one can find the fermionic
part of the Lagrangian:
\begin{eqnarray}
 {\cal L}_F &=& \frac{i}2 \varepsilon^{\mu\nu\rho\sigma}
\bar{\Psi}_{\mu i} \gamma_5 \gamma_{\nu} D_{\rho} \Psi_{\sigma i} +
\frac{i}2 \bar{\lambda}_a^i \hat{D} \lambda_a^i + \frac{i}2
\bar{\chi}^i \hat{D} \chi^i + \frac{i}2 \bar{\Omega}^i \hat{D}
\Omega_i - \nonumber \\
 && - \frac12 \bar{\chi}^i \gamma^{\mu} \gamma^{\nu} \left\{
\partial_{\nu} \varphi \delta_i{}^j - 2e^{\varphi} Q^{a+}_{\nu}
\tau^a{}_i{}^j \right\} \Psi_{\mu j} -  \nonumber \\
 && - \frac12 \bar{\lambda}^i_a \gamma^\mu \gamma^\nu \left\{
(S^+_{\nu} + P_{\nu})_{ab} \tau^b{}_i{}^j + 2e^{\varphi} Q^{a-}_{\nu}
\delta_i{}^j \right\} \Psi_{\mu j} +  \nonumber  \\
 && + \frac{i}2 (S^-_{\mu} + P_{\mu})_{ab} (\bar{\lambda}^i_a
\gamma_{\mu} \lambda^i_b) + 2ie^{\varphi} Q^{a-}_{\mu}
(\bar{\lambda}^i_a \gamma_{\mu} \chi^i) \nonumber \\
 && - (\bar{\lambda}^i_a \gamma^{\mu} \Omega^i) y_{ma} \partial_{\mu}
X^m  - i(\bar{\chi}^i \gamma^{\mu} \Omega^i) e^{\varphi} D_{\mu} X -
\nonumber\\
 && - \frac12 \bar{\Psi}_{\mu i} \gamma^{\nu} \gamma^{\mu} \left\{
e^{\varphi} D_{\nu} X \delta_i{}^j + y_{ma} \partial_{\nu} X^m
\tau^a{}_j{}^i \right\} \Omega^j, \label{e4}
\end{eqnarray}
where D-derivatives for the fermions have the following form:
\begin{equation}
 (D_{\mu})_i{}^j = D_{\mu}^G \delta_i{}^j \pm \frac14
\varepsilon^{abc} (S_{\mu}^- - P_{\mu})_{ab} (\tau^c)_i{}^j +
e^{\varphi} Q^{a+}_{\mu} (\tau^a)_i{}^j 
\end{equation}
Here derivatives of $\Psi_{\mu i}$ and $\eta_i$ have the sign "-" and 
derivatives of $\chi^i$, $\lambda^i_a$ and $\Omega^{iA}$ -- the sign
"+".

  In this, the total Lagrangian is invariant under the following
local $N=2$ supertransformations:
\begin{eqnarray}
 \delta \Psi_{\mu i} &=& 2D_{\mu} \eta_i, \qquad \delta \chi^i = -i
\gamma_{\mu} \left\{ \partial_{\mu} \varphi \delta_i{}^j -
2e^{\varphi} Q^{a+}_{\mu} (\tau^a)_i{}^j \right\} \eta_j,   \nonumber
\\
 \delta \lambda^i_a &=& - i\gamma^{\mu} \left\{ (S^+_{\mu} +
P_{\mu})_{ab} (\tau^b)_i{}^j + 2e^{\varphi} Q^{a-}_{\mu} \delta_i{}^j
\right\} \eta_j, \nonumber \\
 \delta \Omega^i &=& - i\gamma^{\mu} \left\{ e^{\varphi} D_{\mu} X
\delta_i^j + y_{ma} \partial_{\mu} X^m (\tau^a)_i{}^j \right\}
\eta_j,  \nonumber \\ 
 \delta \varphi &=& (\bar{\chi}^i \eta_i), \qquad \quad
\delta y_{ma} = y_{mb} (\bar{\lambda}^i_b (\tau^a)_i{}^j
\eta_j), \label{e6} \\
 \delta \pi^{mn} &=& \frac12 (\bar{\lambda}^i_a (\tau^a)_i{}^j
\eta_j) [y^{ma} y^{nb} - (m \leftrightarrow n)] - \frac12 \left\{ X^m
\delta X^n - (m \leftrightarrow n) \right\}, \nonumber  \\
 \delta \pi_m &=& - e^{-\varphi} y_{ma} [(\bar{\lambda}^i_a \eta_i) +
(\bar{\chi}^i (\tau^a)_i{}^j \eta_j)],  \nonumber   \\
 \delta \l^m &=& \frac14 e^{-\varphi} y^{ma} [(\bar{\lambda}^i_a
\eta_i) - (\bar{\chi}^i (\tau^a)_i{}^j \eta_j)] - \frac12 \pi^{mn}
\delta \pi_n - \frac12 e^{-\varphi} X^m (\bar{\Omega}^i \eta_i) -
\frac14 X^m X^n \delta \pi_n, \nonumber \\
 \delta X^m &=& y^{ma} (\bar{\Omega}^i (\tau^a)_i{}^j \eta_j), \qquad
 \quad \delta X = e^{-\varphi} (\Omega^i \eta_i) - X^m \delta \pi_m.
\nonumber
\end{eqnarray}

  Note, that in what follows by the hidden sector we will mean the
part of this model with the bosonic fields
$(\varphi,Y_{ma},\pi^{mn},\l^m,\pi_n)$  and the fermionic ones
$(\chi^i,\lambda^{ia})$ and the formulae given above where the fields
$(X^A, X^{mA},\Omega^{iA})$ are set to zero.

  Now, let us turn to the $YZ$-sector. The bosonic part of the hidden
sector consists of the field $E_{ma}$, corresponding to the
non-linear $\sigma$-model $GL(4,R)/O(4)$ and antisymmetric tensor
$\Pi^{mn}$. In this formulation the theory has local $O(4)$
invariance, so one can use one of its $O(3)$ subgroup to bring matrix
$E$ to block-triangle form. We will use the following concrete
parameterization:
\begin{equation}
 E = \left( \begin{array}{cc} \sqrt{\Delta} e^{\varphi /2} & 0 \\ -
\frac12 \sqrt{\Delta} e^{\varphi /2} \varepsilon_{mnk} \pi^{nk} &
\frac1{\sqrt{\Delta}} e^{\varphi /2} y_{ma} \end{array} \right),
\qquad \Pi = \left( \begin{array}{cc} 0 & -2 \l^m \\ 2 \l^n & \frac12
\varepsilon^{mnk} \pi_k \end{array} \right),
\end{equation}
where now $m,a = 1,2,3$, $\Delta = det(y_{ma})$. In this, the hidden
sector takes exactly the same form as in the formulas given above (of
course, in the absence of the fields $X$, $X^m$ and $\Omega^i$).
Scalar fields of $Y$ and $Z$ multiplets, which are now
$(Y,Y_m)^{\dot{A}}$ and $(Z,Z^m)^{\ddot{A}})$, give the following
contribution to the bosonic Lagrangian:
\begin{eqnarray}
{\cal L}^B &=& \frac{e^{\varphi}}{2\Delta} (\partial_{\mu} Y)^2 + 
e^{\varphi} \frac{\Delta}{2} (D_{\mu}Y_m) (D_{\mu}Y_n) g^{mn} +
\nonumber \\
 && + e^{\varphi} \frac{\Delta}2 (D_{\mu}Z)^2 +
\frac{e^{\varphi}}{2\Delta} (D_{\mu}Z^m) (D_{\mu}Z^n) g_{mn},
\label{e8}
\end{eqnarray}
where 
\begin{eqnarray}
 D_{\mu} Y_m &=& \partial_{\mu} Y_m - \varepsilon_{mnk} \pi^{nk}
\partial_{\mu} Y,  \qquad D_{\mu} Z^m = \partial_{\mu} Z^m, \nonumber
\\
 D_{\mu} Z &=& \partial_{\mu} Z + \varepsilon_{mnk} \pi^{mn}
\partial_{\mu} Z^k.
\end{eqnarray}
At the same time our definitions for $U_{\mu m}$ and $L_{\mu}^m$
change to:
\begin{eqnarray}
 U_{\mu m} &=& \partial_\mu \pi_m + (Y_m
 \stackrel{\leftrightarrow}{\partial_{\mu}} Y) - \frac12
\varepsilon_{mnk} (Z^n\stackrel{\leftrightarrow}
{\partial_{\mu}}Z^k), \label{e10} \\
 L_{\mu}^m &=&  \partial_{\mu} \l^m + \frac12 \pi^{mn} U_{\mu n} - 
\frac18 \varepsilon^{mnk}
(Y_n\stackrel{\leftrightarrow}{\partial_{\mu}}
Y_k) + \frac14 (Z^m\stackrel{\leftrightarrow}{\partial_{\mu}} Z).
\label{e11}
\end{eqnarray}

  The resulting bosonic Lagrangian, besides the usual $GL(3,R)$,
scale transformations and trivial translations for the fields $\l^m$
and $\pi_m$, is invariant under the following global transformations:
\begin{eqnarray}
 \delta \pi^{mn} &=& - 2 \Lambda^{mn}, \quad \delta \l^m =
\Lambda^{mn} \pi_n, \quad \delta Y_m = - \varepsilon_{mnk}
\Lambda^{nk} Y, \quad \delta Z = \varepsilon_{mnk} \Lambda^{m} Z^k,
\nonumber \\
 \delta Y &=& \xi, \quad \delta \pi_m = \xi Y_m, \qquad \delta Y_m =
\xi_m, \quad \delta \l^m = \frac14 \varepsilon^{mnk} \xi_n Y_k, \quad
\delta \pi_m = - \xi_m Y, \\
 \delta Z &=& \eta^m, \quad \delta \l^m = \frac14 \eta Z^m, \qquad
\delta Z^m = \eta^m, \quad \delta \l^m = - \frac14 \eta^m Z, \quad
\delta \pi_m = \varepsilon_{mnk} \eta^n Z^k. \nonumber
\end{eqnarray}

The fermionic Lagrangian, containing the fields of the Y- and
Z-multiplets, has the following form:
\begin{eqnarray}
 {\cal L}^F &=& \frac{i}2 \bar{\Lambda}^i \hat{D} \Lambda^i - \frac12
e^{\varphi/2} \bar{\Psi}_{\mu i} \gamma^{\nu} \gamma^{\mu}
(V_{\nu})_i{}^j \Lambda^j + \nonumber  \\
 && + \frac{i}2 \bar{\Omega}^i \hat{D} \Omega^i - \frac12 e^{\varphi
/2} \bar{\Psi}_{\mu i} \gamma^{\nu} \gamma^{\mu} (W_{\nu})_i{}^j
\Sigma^j - \nonumber   \\
 && - \frac{i}2 e^{\varphi /2} \bar{\chi}^i \gamma^{\mu}
(V_{\mu})_i{}^j \Lambda^j - \frac{i}2 e^{\varphi /2} \bar{\chi}^i
\gamma^{\mu} (W_{\mu})_i{}^j \Sigma^j - \nonumber   \\
 && - \frac{i}2 e^{\varphi /2} \bar{\lambda}^i_a \gamma^{\mu}
(V_{\mu})_j{}^i (\tau^a)_k{}^j \Lambda^k + \frac{i}2 e^{\varphi /2}
\bar{\lambda}^i_a \gamma^{\mu} (W_{\mu})_j{}^i (\tau^a)_k{}^j
\Sigma^k, \label{e13} 
\end{eqnarray}
where $D_\mu$ for the $\Lambda^i$ and $\Sigma^i$ are the same as one
for $\chi^i$ and
\begin{eqnarray}
 (V_{\mu})_i{}^j &=& \frac1{\sqrt{\Delta}} \partial_{\nu} Y
\delta_i{}^j + \sqrt{\Delta} D_{\nu} Y_m y^{ma} (\tau^a)_i{}^j,
\nonumber  \\
 (W_{\mu})_i{}^j &=& \sqrt{\Delta} D_{\mu} Z \delta_i{}^j +
\frac1{\sqrt{\Delta}} D_{\mu} Z^m y_{ma} (\tau^a)_i{}^j,
\end{eqnarray}
while the supertransformations for the fields of $Y$ and $Z$
hypermultiplets are the following:
\begin{eqnarray}
 \delta Y &=& \sqrt{\Delta} e^{-\varphi/2} (\bar{\Lambda}^i \eta_i),
\qquad \delta Z^m = \sqrt{\Delta} e^{-\varphi/2} y^{ma}
(\bar{\Sigma}^i (\tau^a)_i{}^j \eta_j),  \nonumber   \\
 \delta Z &=& \frac1{\sqrt{\Delta}} e^{-\varphi/2} (\bar{\Sigma}^i
\eta_i) - \varepsilon_{mnk} \pi^{mn} \delta Z^k,  \label{e15} \\
 \delta Y_m &=& \frac1{\sqrt{\Delta}} e^{-\varphi/2} y_{ma}
(\bar{\Lambda}^i (\tau^a)_i{}^j \eta_j) + \varepsilon_{mnk} \pi^{mn}
\delta Y,  \nonumber \\
 \delta \Lambda^i &=& - i \gamma^{\mu} (V_{\mu})_i{}^j \eta_j, \qquad
\delta \Sigma^i = - i \gamma^{\mu} (W_{\mu})_i{}^j \eta_j. \nonumber
\end{eqnarray}

Besides, some new terms in the supertransformation laws of the fields 
$\pi_m$ and $l^m$ appear:
\begin{eqnarray}
 \delta'l^m &=& \frac14 \varepsilon^{mnk} Y_n \delta Y_k + \frac14 Z
\delta Z^m - \frac14 Z^m \delta Z,   \nonumber   \\
 \delta'\pi_m &=& Y \delta Y_m - Y_m \delta Y + \varepsilon_{mnk} Z^n
\delta Z^k.
\end{eqnarray}

  Now we are ready to construct a model, containing both the X-sector
and the YZ-sector, described above. The method is, starting 
from the X-sector, to add Y and Z-multiplets  by the use of the usual
Noether procedure, extending $U_{\mu m}$ and $L_{\mu}^m$ as in
formulae (\ref{e10}), (\ref{e11}). As a result, all the terms, which
are present in the pure YZ-model, appear as well as some new,
"crossing" terms, containing fermions from both X- and
Y,Z-multiplets. In this, we have to introduce constant matrices
$\Gamma^{A\dot{A}\ddot{A}}$, carrying all three kinds of the indices
in order to connect the fields of different multiplets. The bosonic
symmetries and the supersymmetry impose certain constraints on these
matrices --- they turn out to be $\gamma$-matrices for the $O(p)$
group. The calculations are tedious and formidable, but, at least
partly, interesting results can be obtained by means of the
symmetry considerations.

  All the changes in the bosonic sector of unified XYZ-model as
compared with pure X and YZ-sectors can be rather easily seen by
exploring the symmetries of the bosonic Lagrangian. For example,
consider one of the bosonic symmetries of the X-model that we denoted
$\zeta^m$ (first line in (\ref{e3})). The combination $D_{\mu}X^A$
(\ref{e2}) is invariant under this transformation. But, as it has
been shown above, adding Y and Z-multiplets, we have to "extend"
$\partial_{\mu}\pi_m$ to $U_{\mu m}$ according to (\ref{e10}). It is
easy to check, that the combination $D_{\mu}X^A$ with such $U_{\mu
m}$ is already noninvariant under the $\zeta^m$ transformations and
in order to restore this symmetry we have to "extend" "covariant
derivative" $D_{\mu}X$:
\begin{equation}
 D_\mu X^A \rightarrow D_\mu X^A + \frac12 \Gamma^{A\dot{A}\ddot{A}}
[(Y^{\dot{A}} \stackrel{\leftrightarrow}{\partial_\mu} Z^{\ddot{A}})
+ (Y^{\dot{A}}_m \stackrel{\leftrightarrow}{\partial_\mu}
Z^{m\ddot{A}})]. \label{e18}
\end{equation}
In this, the fields Y and Z are transformed under
$\zeta^m$-transformations according to the following formulas:
\begin{eqnarray}
 \delta Y^{\dot{A}} &=& 0, \qquad  \qquad
\delta Y_m^{\dot{A}} = \Gamma^{A\dot{A}\ddot{A}} \varepsilon_{mnk}
\zeta^{nA} Z^{k\ddot{A}},   \nonumber   \\
 \delta Z^{\ddot{A}} &=& \Gamma^{A\dot{A}\ddot{A}} \zeta^{mA}
Y_m^{\dot{A}},  \qquad \delta Z^{m\ddot{A}} = -
\Gamma^{A\dot{A}\ddot{A}} \zeta^{mA} Y^{\dot{A}}
\end{eqnarray}
that leaves "extended" $U_{\mu m}$ (\ref{e10}) invariant. 

The requirement of the invariance of $D_{\mu}X$ under
$\zeta^m$-transformations as well as the requirement of the closure
of the corresponding algebra leads to the constraint on the
$\Gamma$-matrices:
\begin{equation}
 \Gamma^{A\dot{A}\ddot{A}} \Gamma^{B\dot{B}\ddot{A}} +
(A\leftrightarrow B) = 2 \delta^{AB} \delta^{\dot{A}\dot{B}},
\end{equation}
i. e., just as it should be, they are the $\gamma$-matrices for the
$O(p)$ group. The commutator of the two $\zeta^m$-transformations
gives a transformation, which is a part of the O(4,4) group (second
line in (\ref{e3})):
\begin{eqnarray}
 \delta \pi^{mn} &=& - 2 \Lambda^{mn},  \qquad
\delta l^m = \Lambda^{mn} \pi_n,   \nonumber   \\
 \delta Y_m &=& - 2 \Lambda_m Y,  \qquad  
\delta Z = 2 \Lambda_m Z^m,
\end{eqnarray}
where $\Lambda^{mn} \sim [\zeta_1^{mA} \zeta_2^{nA} -
(m\leftrightarrow  n)]$, $\Lambda_m = \varepsilon^{mnk} \Lambda^{nk}$
and all other fields are inert under this transformation.

   Due to the fact, that, for example, field $Z^m$ transforms
nontrivially under the $\zeta^m$-transformation, the derivative
$\partial_{\mu}Z^m$ is noninvariant under it. In order to restore the
invariance we have to "extend" this derivative to
\begin{equation}
 D_\mu Z^{m\ddot{A}} = \partial_\mu Z^{m\ddot{A}} + 
\Gamma^{A\dot{A}\ddot{A}} X^{mA} \partial_\mu Y^{\dot{A}}.
\end{equation}
A bit more formidably, but by means of the same considerations, one
can obtain the corresponding expressions, invariant under the
$\zeta^m$-transformations, for the "extended" derivatives of the
$Z^{\ddot{A}}$ and $Y_m^{\dot{A}}$ fields:
\begin{eqnarray}
 D_\mu Y_m^{\dot{A}} &=& \partial_\mu Y_m^{\dot{A}} -
\varepsilon_{mnk} [\pi^{nk} \delta^{\dot{A}\dot{B}} + \frac12 X^{nA}
X^{kB} (\Sigma^{AB})^{\dot{A}\dot{B}}] \partial_\mu Y^{\dot{B}} -
\nonumber  \\
 && - \varepsilon_{mnk} X^{nA} \Gamma^{A\dot{A}\ddot{A}} \partial_\mu
Z^{k\ddot{A}}, \label{e23} \\
 D_\mu Z^{\ddot{A}} &=& \partial_\mu Z^{\ddot{A}} + \varepsilon_{mnk}
[\pi^{mn} \delta^{\ddot{A}\ddot{B}} + \frac12 X^{mA} X^{nB}
(\Sigma^{AB})^{\ddot{A}\ddot{B}}] \partial_\mu Z^{k\ddot{B}} -
\nonumber \\
 && - X^{mA} \Gamma^{A\dot{A}\ddot{A}} (\partial_\mu Y_m^{\dot{A}} -
\varepsilon_{mnk} \pi^{nk} \partial_\mu Y^{\dot{A}}) -  \nonumber  \\
 && - \frac16 \varepsilon_{mnk} X^{mA} X^{nB} X^{kC}
(\Gamma^{ABC})^{\dot{A}\ddot{A}} \partial_\mu Y^{\dot{A}},
\nonumber
\end{eqnarray}
where $\Sigma^{AB} = \frac12 (\Gamma^A \Gamma^B - \Gamma^B \Gamma^A)$
and $(\Gamma^{ABC})^{\dot{A}\ddot{A}} = \frac16
[\Gamma^{A\dot{A}\ddot{B}} \Gamma^{B\dot{B}\ddot{B}}
\Gamma^{C\dot{B}\ddot{A}} + (ABC-cycle)]$. The derivative
$\partial_{\mu}Y^{\dot{A}}$ does not change its form because the
field $Y^{\dot{A}}$ is inert under $\zeta^m$-transformation.

  The total bosonic Lagrangian of the $V(p,q)$-model is just the sum
of the bosonic Lagrangians (\ref{e1}) and (\ref{e8}), where now
\begin{eqnarray}
 L_\mu^m &=& \partial_\mu l^m + \frac12 \pi^{mn} U_{\mu n} + \frac12
X^{mA} D_\mu X^A - \frac14 X^{mA} X^{nA} U_{\mu n} -  \nonumber \\
 && - \frac18 \varepsilon^{mnk} (Y_n^{\dot{A}}
\stackrel{\leftrightarrow} {\partial_\mu} Y_k^{\dot{A}}) + \frac14
(Z^{m\ddot{A}} \stackrel{\leftrightarrow}{\partial_\mu} Z^{\ddot{A}})
\end{eqnarray}
and all other covariant objects are defined in (\ref{e2}),
(\ref{e10}), (\ref{e18}) and (\ref{e23}).

 The corresponding fermionic Lagrangian is the sum of the fermionic
Lagrangians (\ref{e4}) and (\ref{e13}) with some additional
"crossing" terms, which have the following form:
\begin{eqnarray}
 \Delta {\cal L}^F = &-& \frac{i}2 e^{\varphi/2}
\Gamma^{A\dot{A}\ddot{A}} \bar{\Omega}^{iA} \gamma^\mu (V_\mu)_j{}^i
\Sigma^{j\ddot{A}} + \frac{i}2 e^{\varphi/2}
\Gamma^{A\dot{A}\ddot{A}} \bar{\Omega}^{iA} \gamma^\mu (W_\mu)_j{}^i
\Lambda^{j\dot{A}} - \nonumber   \\
 &-& \frac{i}2 \Gamma^{A\dot{A}\ddot{A}} \bar{\Lambda}^{i\dot{A}}
\gamma^\mu \left\{ e^{\varphi} D_\mu X^A \delta_j{}^i - \partial_\mu
X^{mA} y_{ma} (\tau^a)_j{}^i \right\} \Sigma^{j\ddot{A}}.
\end{eqnarray}
In this, the supertransformations for the fermionic fields are the
same as in formulas (\ref{e6}) and (\ref{e15}) taking into
account the changes in the expressions such as $D_\mu X$, $U_{\mu m}$
and so on.

   The supertransformations of the bosonic fields are the following:
\begin{eqnarray}
 \delta \pi^{mn} &=& \frac12 (\bar{\lambda}^i_a (\tau^a)_i{}^j
\eta_j) [y^{ma} y^{nb} - (m \leftrightarrow n)] - \frac12 [ X^{mA}
\delta X^{nA} - (m \leftrightarrow n)],    \nonumber   \\
\delta Y^{\dot{A}} &=& \sqrt{\Delta} e^{-\varphi/2}
(\bar{\Lambda}^{i\dot{A}} \eta_i)  \qquad   \delta X^{mA} = y^{mA}
(\bar{\Omega}^{iA} (\tau^a)_i{}^j \eta_j),  \nonumber  \\   
\delta Z^{m\ddot{A}} &=& \sqrt{\Delta} e^{-\varphi/2} y^{ma}
(\bar{\Sigma}^ {i\ddot{A}} (\tau^a)_i{}^j \eta_j) - X^{mA}
\Gamma^{A\dot{A}\ddot{A}} \delta Y^{\dot{A}},  \nonumber   \\
\delta Y_{m\dot{A}} &=& \frac1{\sqrt{\Delta}} e^{-\varphi/2} y_{ma}
(\bar {\Lambda}^{i\dot{A}} (\tau^a)_i{}^j \eta_j) + \varepsilon_{mnk}
X^{nA} \Gamma^{A\dot{A}\ddot{A}} \delta Z^{k\ddot{A}} + \nonumber  \\
 && + \varepsilon_{mnk} [\pi^{mn} \delta^{\dot{A}\dot{B}} + \frac12
X^{nA} X^{kB} (\Sigma^{AB})^{\dot{A}\dot{B}}] \delta Y^{\dot{B}},
\nonumber   \\
\delta Z^{\ddot{A}} &=& \frac1{\sqrt{\Delta}} e^{-\varphi/2}
(\bar{\Sigma}^{i\ddot{A}} \eta_i) + X^{mA} \Gamma^{A\dot{A}\ddot{A}}
\delta Y_m^{\dot{A}} - \nonumber   \\
 && - \varepsilon_{mnk} [ \pi^{mn} \delta^{\ddot{A}\ddot{B}} +
\frac12 X^{mA} X^{nB} (\Sigma^{AB})^{\ddot{A}\ddot{B}}] \delta
Z^{k\ddot{B}} -  \nonumber  \\
&& - \varepsilon_{mnk} X^{mA} \Gamma^{A\dot{A}\ddot{B}} [ \pi^{nk}
\delta^{\ddot{A}\ddot{B}} - \frac16 X^{nB} X^{kC}
(\Sigma^{BC})^{\ddot{B}\ddot{A}}] \delta Y^{\dot{A}},   \nonumber  \\
\delta \pi_m &=& - e^{-\varphi} y_{ma} [ (\bar{\lambda}^i_a \eta_i) +
(\bar{\chi}^i (\tau^a)_i{}^j \eta_j) ] + Y^{\dot{A}} \delta
Y_m^{\dot{A}} - Y_m^{\dot{A}} \delta Y^{\dot{A}} + \varepsilon_{mnk}
Z^{n\ddot{A}} \delta Z^{k\ddot{A}},  \nonumber  \\
\delta X^A &=& e^{-\varphi} (\Omega^{iA} \eta_i) - X^{mA} \delta_0
\pi_m - \nonumber  \\
 && - \frac12 \Gamma^{A\dot{A}\ddot{A}} \left\{ Y^{\dot A} \delta
Z^{\ddot{A}} - Z^{\ddot{A}} \delta Y^{\dot{A}} Y_m^{\dot A} \delta
Z^{m\ddot{A}} - Z^{m\ddot{A}} \delta Y_m^{\dot{A}}\right\}, \nonumber
\\
\delta \l^m &=& \frac14 e^{-\varphi} y^{ma} [ (\bar{\lambda}^i_a
\eta_i) - (\bar{\chi}^i (\tau^a)_i{}^j \eta_j) ] - \frac12 \pi^{mn}
\delta_0 \pi_n - \frac12 e^{-\varphi} X^{mA} (\bar{\Omega}^{iA}
\eta_i)   \nonumber \\
 && - \frac14 X^m X^n \delta_0 \pi_n + \frac14 \varepsilon^{mnk}
Y_n^{\dot{A}} \delta Y_k^{\dot{A}} + \frac14 Z^{\ddot{A}} \delta
Z^{m\ddot{A}} - \frac14 Z^{m\ddot{A}} \delta Z^{\ddot{A}},\label{40}
\end{eqnarray}
where
$\delta_0 \pi_m = - e^{-\varphi} y_{ma} [ (\bar{\lambda}^i_a \eta_i)
+ (\bar{\chi}^i (\tau^a)_i{}^j \eta_j)]$. As it has already been
said, all these formulas can be obtained by the use of the
straightforward Noether procedure.

\section*{Conclusion}
  So, we have constructed the Lagrangian and supertransformations for
the two general types of $N=2$ supergravity models, based on the
nonsymmetric quaternionic manifolds. In the following paper we will
consider the gauge interactions which are possible in such models,
in-particular, the ones that lead to the spontaneous supersymmetry
breaking.

\vspace{0.3in}
{\large \bf Acknowledgements}
\vspace{0.2in}

The work was supported by the International Science Foundation and
Russian Government grant RMP300 and by the Russian Foundation for
Fundamental Research grant 95-02-06312.

\newpage

\appendix
\section{}

  Let us first give the explicit expressions for the $T_{\pm\pm\pm}$
generators in terms of $\Lambda^{\pm}$ combinations determined in
Section 1:
\begin{eqnarray}
 T_{+++} &=& \Omega_1 + \Omega_2 + \Omega_3 + \Omega_4 -
\Omega_5 - \Omega_6 - \Omega_7 - \Omega_8, \nonumber \\
 T_{-++} &=& \Omega_1 + \Omega_2 - \Omega_3 - \Omega_4 +
\Omega_5 + \Omega_6 - \Omega_7 - \Omega_8, \nonumber \\
 T_{+-+} &=& \Omega_1 - \Omega_2 + \Omega_3 - \Omega_4 +
\Omega_5 - \Omega_6 + \Omega_7 - \Omega_8, \nonumber \\
 T_{--+} &=& \Omega_1 - \Omega_2 - \Omega_3 + \Omega_4 -
\Omega_5 + \Omega_6 + \Omega_7 - \Omega_8, \nonumber \\
 T_{++-} &=& \Omega_1 - \Omega_2 - \Omega_3 + \Omega_4 +
\Omega_5 - \Omega_6 - \Omega_7 + \Omega_8,  \\
 T_{-+-} &=& \Omega_1 - \Omega_2 + \Omega_3 - \Omega_4 -
\Omega_5 + \Omega_6 - \Omega_7 + \Omega_8, \nonumber \\
 T_{+--} &=& \Omega_1 + \Omega_2 - \Omega_3 - \Omega_4 -
\Omega_5 - \Omega_6 + \Omega_7 + \Omega_8, \nonumber \\
 T_{---} &=& \Omega_1 + \Omega_2 + \Omega_3 + \Omega_4 +
\Omega_5 + \Omega_6 + \Omega_7 + \Omega_8, \nonumber
\end{eqnarray}
where all $\Omega$ stand for $\Omega^+$. The expressions for the
$\hat{T}_{\pm\pm\pm}$ are similar to the ones given above.

 Now we give the commutation relations for the $F_4$ generators in
our multispinor basis. They look like:
\begin{equation}
  [ \Lambda_{i\alpha}, \Lambda^{j\beta} ] = ( t_i{}^j
\delta_\alpha{}^\beta + \delta_i{}^j t_\alpha{}^\beta ), \qquad
[ \Lambda_{\alpha\dot{\alpha}}, \Lambda^{\beta\dot{\beta}} ] = - (
t_\alpha{}^\beta \delta_{\dot{\alpha}}{}^{\dot{\beta}} +
\delta_\alpha{}^\beta t_{\dot{\alpha}}{}^{\dot{\beta}} ), \nonumber
\end{equation}
\begin{equation}
  [ \Lambda_{i\alpha}, \Lambda_{j\dot{\alpha}} ] = \varepsilon_{ij}
\Lambda_{\alpha\dot{\alpha}}, \qquad [ \Lambda_{i\alpha},
\Lambda_{\beta\dot{\alpha}} ] = - \varepsilon_{\alpha\beta}
\Lambda_{i\dot{\alpha}},
\end{equation}
\begin{equation}
  [ \Lambda_{i\alpha}, \Lambda_{\dot{\alpha}\ddot{\alpha}} ] = [
\Lambda_{i\dot{\alpha}}, \Lambda_{\alpha\ddot{\alpha}} ] = [
\Lambda_{i\ddot{\alpha}}, \Lambda_{\alpha\dot{\alpha}} ] =
T_{i\alpha\dot{\alpha}\ddot{\alpha}}, \nonumber
\end{equation}
\begin{equation}
  [ T_{i\alpha\dot{\alpha}\ddot{\alpha}}, \Lambda^{j\beta} ] = -
\frac1{\sqrt{2}} \delta_i{}^j \delta_\alpha{}^\beta
\Lambda_{\dot{\alpha}\ddot{\alpha}}, \nonumber
\end{equation}
and a lot of similar ones for other positions of indices.

  The following combinations of these generators are the
eighenvectors for the $h_{1,2,3,4}$:
\begin{eqnarray}
 X_\pm &=& \Lambda_{0012} \pm \Lambda_{0021} \pm \Lambda_{1200} +
\Lambda_{2100}, \nonumber \\
 \hat{X}_\pm &=& \Lambda_{0012} \mp \Lambda_{0021} \pm \Lambda_{1200}
- \Lambda_{2100}, \nonumber \\
 Y_\pm &=& \Lambda_{0102} \pm \Lambda_{0201} \pm \Lambda_{1020} +
\Lambda_{2010}, \nonumber \\
 \hat{Y}_\pm &=& \Lambda_{0102} \mp \Lambda_{0201} \pm \Lambda_{1020}
- \Lambda_{2010}, \\
 Z_\pm &=& \Lambda_{0120} \pm \Lambda_{0210} \pm \Lambda_{1002} +
\Lambda_{2001}, \nonumber \\
 \hat{Z}_\pm &=& \Lambda_{0120} \mp \Lambda_{0210} \pm \Lambda_{1002}
- \Lambda_{2001}, \nonumber
\end{eqnarray}
where, for example, $\Lambda_{1200}$ stands for $\Lambda_{i\alpha}$,
$i=1,\alpha=2$ as well as another twelve combinations which we
denote as $\tilde{X}_\pm$, $\hat{\tilde{X}}_\pm$, $\tilde{Y}_\pm$,
$\hat{\tilde{Y}}_\pm$, $\tilde{Z}_\pm$ and $\hat{\tilde{Z}}_\pm$.

\end{document}